\documentclass[conference]{IEEEtran}
\IEEEoverridecommandlockouts
\usepackage{cite}
\usepackage{amsmath,amssymb,amsfonts}
\usepackage{algorithmic}
\usepackage{graphicx}
\usepackage{textcomp}
\usepackage{xcolor}
\def\BibTeX{{\rm B\kern-.05em{\sc i\kern-.025em b}\kern-.08em
    T\kern-.1667em\lower.7ex\hbox{E}\kern-.125emX}}
\begin{document}

\title{A Lightweight Dual-Factor Acoustic Authentication System via Cascaded GMM-DTW Architecture for Edge Computing\\
}

\author{\IEEEauthorblockN{1\textsuperscript{st} Yutong Zhang}
\IEEEauthorblockA{\textit{The School of Computer Science} \\
\textit{Northeast Electric Power University}\\
Jilin 132012, China \\
2023303030110@neepu.edu.cn}
}

\maketitle
\begin{abstract}
This paper presents a lightweight, cascaded GMM-DTW dual-factor voice lock system for resource-constrained edge environments. By utilizing a shared MFCC feature space, the framework implements a sequential defense mechanism combining GMM speaker screening and DTW passphrase verification. To counter presentation threats without extra hardware, a dynamic joint absolute-relative margin constraint is integrated into the GMM classification space, limiting the physical imposter and high-fidelity replay attack False Acceptance Rates (FAR) to \textbf{2.73\%} and \textbf{6.67\%}, respectively, with a legitimate False Rejection Rate (FRR) of \textbf{16.67\%}. Due to Sakoe-Chiba window optimization, the global end-to-end processing latency under temporal stress is rigidly bounded at \textbf{9.82~ms} on a single-core CPU, comprising 1.51~ms for feature extraction, 0.54~ms for GMM scoring, and 7.77~ms for worst-case DTW matching. These empirical benchmarks demonstrate the viability of white-box acoustic cascades for secure, deterministic real-time deployment on low-power edge nodes.
\end{abstract}


\section{Introduction}
With the proliferation of the Internet of Things (IoT) and edge computing in smart homes and access control, localized biometric identification is vital for physical security \cite{lien2023challenges}. Voice authentication is particularly attractive due to its non-contact acquisition, low cost, and high user acceptance \cite{lien2023challenges, koffi2023voice}. Although early text-dependent template matching was computationally efficient, it suffered from ambient noise and DC offsets \cite{zheng2023text}. Recent deep learning architectures (e.g., CNNs\cite{paul2025isolated}, RNNs\cite{sharrab2025advancements, wu2020multi}, and Transformers\cite{sharrab2025advancements, thorbecke2024fast}) have significantly advanced Automatic Speech Recognition (ASR) and Voiceprint Speaker Recognition (VSR), establishing a modern paradigm for feature extraction rooted in statistical probability and auditory perception, such as combining Gammatone frequency cepstral coefficients (GFCCs) with GMM classifiers \cite{pr2023research, krobba2023novel}.

Despite cloud-level success, deploying deep neural networks onto resource-constrained edge devices exposes two physical bottlenecks \cite{raza2025lightweight}. First, their massive parameter sizes and floating-point operations (FLOPs) require GPU acceleration, causing prohibitive inference latency and energy consumption on CPU-only nodes \cite{more2023lightweight, zhang2025sparoa}. Second, relying on a single voiceprint factor (verifying only who is speaking) introduces severe security vulnerabilities \cite{soni2016text}. Because traditional algorithms exhibit overlapping decision boundaries in multi-class feature spaces \cite{lee2018overlap}, single-factor architectures lack robustness against unknown speaker spoofing or authorized replay attacks \cite{vo2026charvoc}, elevating false-acceptance risks.

To circumvent these computational and security constraints, this paper dispenses with energy-intensive, black-box deep learning. Instead, we propose a lightweight, edge-oriented "voiceprint-password" dual-factor authentication system built entirely on classical acoustic pipelines\cite{gonzalez2025deploying} and statistical machine learning \cite{yang2026low}. Our core contributions are threefold:

\begin{itemize}
  \item \textbf{Cascaded Liveness Defense:} A non-linear cascade combining a GMM coarse filter with a newly proposed Dynamic Likelihood Space Constriction (DLSC) mechanism and a DTW verifier, enabling physical-level authentication and an implicit anti-replay barrier under joint absolute-relative margin constraints.
  \item \textbf{Acoustic Feature Reuse:} A shared feature space driven by Mel-Frequency Cepstral Coefficients (MFCC) extraction, ensuring high feature representation density while maximizing computational efficiency through downstream classifier reuse.
  \item \textbf{High Efficiency and Determinism:} A deterministic end-to-end processing latency bounded at \textbf{9.82~ms} on a pure single-core CPU, ensuring bounded real-time execution within 10~ms even under extreme linguistic variations.
\end{itemize}

\section{System Framework and Feature Decomposition}

\subsection{Overall Cascaded Architecture Workflow}

\begin{figure}[!t]
    \centering
    \includegraphics[width=\columnwidth]{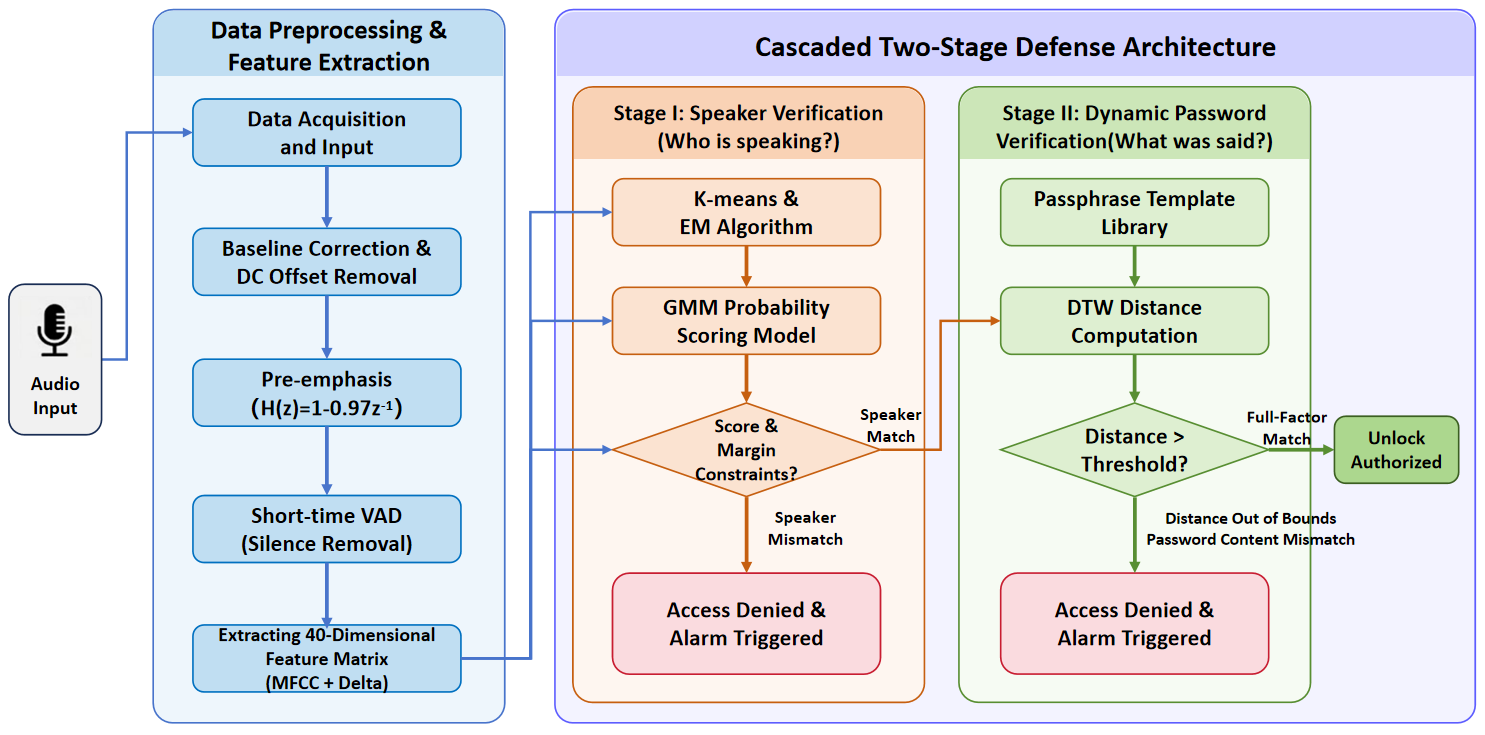}
    \caption{System workflow diagram.}
    \label{fig1}
\end{figure}

The proposed dual-factor acoustic authentication system deploys a two-stage cascaded defense architecture (Fig. \ref{fig1}), establishing a sequential speaker voiceprint probabilistic lock and a textual password content lock.

Upon signal input, a shared front-end pipeline performs standardized pre-processing and feature extraction. The resulting feature matrix is first evaluated by a trained multi-speaker Gaussian Mixture Model (GMM) network; if the log-likelihood score falls below an adaptive voiceprint threshold, the system intercepts the signal as an unknown speaker spoofing attack. Otherwise, this matrix is immediately reused by the Dynamic Time Warping (DTW) engine for non-linear sequence alignment against an isolated-word template library. Access is granted exclusively when the cumulative warping distance satisfies the password threshold and the spoken passphrase text is verified as correct.

\subsection{Baseline Correction and Endpoint Pre-processing}

To eliminate low-frequency interference introduced at the electrophysical layer of acquisition devices, the front-end pipeline first performs first-order detrending and DC offset correction. Defining the input discrete speech signal as $x(n)$ with a sampling rate of $f_s$, a first-order polynomial is fitted via least-squares estimation to model the linear trend:

\begin{equation}
x_{\text{trend}}(n) = an + b
\end{equation}

Subtracting this trend forces the statistical expectation of the corrected signal to return to a zero-mean baseline:

\begin{equation}
y(n) = x(n) - x_{\text{trend}}(n)
\end{equation}

Subsequently, the signal passes through a time-domain first-order high-pass filter for pre-emphasis\cite{piralideh2024zero}, compensating for the natural glottal excitation attenuation ($-6\text{ dB/oct}$) in high-frequency speech bands. Its system function is expressed as:

\begin{equation}
H(z) = 1 - \alpha z^{-1}
\end{equation}

Where the pre-emphasis coefficient $\alpha$ is set to $0.97$. The pre-emphasized signal is then routed to a dual-threshold time-domain Voice Activity Detection (VAD) module based on Short-Time Energy (STE) and Short-Time Zero-Crossing Rate (ZCR)\cite{yiming2015voice}. This module precisely truncates silent segments and ambient background noise from both ends, extracting the effective high-SNR speech core.

\subsection{High-Order Auditory Feature Extraction}

The system reconstructs the higher-order acoustic feature matrix based on the nonlinear auditory perception characteristics of the human ear. The discrete power spectrum, obtained via windowed framing (frame length: 200 points, frame shift: 100 points, using a Hamming window\cite{barai2023empirical}), is projected onto a 24-order overlapping Mel-filter bank. The mapping between the Mel scale and standard Hertz frequency follows:
\begin{equation}
M(f) = 1127 \ln \left( 1 + \frac{f}{700} \right)
\end{equation}

By computing the weighted energy sum within each bandpass channel, the high-dimensional linear power spectrum is efficiently compressed into a 20-dimensional Mel-logarithmic energy spectrum. Subsequently, the Discrete Cosine Transform (DCT) is applied to decorrelate the inter-channel coefficients, extracting the top 20-dimensional static Mel-Frequency Cepstral Coefficients (MFCCs)\cite{zhang2013application}.

To further capture the temporal dynamic contextual properties of the speech signal, a first-order time derivative (Delta) operation\cite{ali2025audio} is performed on the static feature matrix:
\begin{equation}
\Delta C_m(t) = \frac{\sum_{k=1}^{N} k [C_m(t+k) - C_m(t-k)]}{2 \sum_{k=1}^{N} k^2}
\end{equation}

Where the time window half-length $N$ is set to 2. Finally, the 20-dimensional static MFCCs and the 20-dimensional dynamic delta MFCCs are concatenated along the temporal axis. This reconstructs a time-varying, high-order acoustic feature matrix with a single-frame dimension of 40, which serves as the core underlying representation reused by the downstream cascaded classifier.

\section{Dual-Factor Authentication Classifier Design}
\subsection{First Factor: GMM-Based Speaker Voiceprint Probabilistic Space Modeling}

The first stage deploys a Gaussian Mixture Model (GMM)\cite{saakshara2020speaker} to capture the statistical voiceprint distribution. For a 40-dimensional feature vector $\mathbf{x}$ from a valid speech frame, the probability density function for a registered speaker $S_k$ is a weighted combination of $M$ Gaussian components:
\begin{equation}
P(\mathbf{x} | S_{k}) = \sum_{i=1}^{M} w_{i} \cdot \mathcal{N}(\mathbf{x}; \boldsymbol{\mu}_{i}, \mathbf{\Sigma}_{i})
\end{equation}
Where $w_i$ is the mixture weight satisfying $\sum_{i=1}^{M} w_i = 1$, and $\mathcal{N}(\mathbf{x}; \boldsymbol{\mu}_i, \mathbf{\Sigma}_i)$ is the multivariate Gaussian distribution with mean $\boldsymbol{\mu}_i$ and covariance $\mathbf{\Sigma}_i$. To minimize edge storage and boost inference efficiency, we restrict $\mathbf{\Sigma}_i$ to a diagonal form ($\text{covariance\_type='diag'}$) and set $M = 4$.

Initialized via K-means, the model parameters $\{w_i, \boldsymbol{\mu}_i, \mathbf{\Sigma}_i\}$ are iteratively optimized via the Expectation-Maximization (EM) algorithm to maximize the accumulated log-likelihood\cite{blomer2016adaptive}. During testing, the average log-likelihood score of a sequence $\mathbf{X} = \{\mathbf{x}_1, \mathbf{x}_2, \dots, \mathbf{x}_T\}$ under model $S_k$ is defined as:
\begin{equation}
\Lambda(\mathbf{X} | S_k) = \frac{1}{T} \sum_{t=1}^{T} \ln P(\mathbf{x}_t | S_k)
\end{equation}

The predicted speaker label is retrieved via $\hat{k} = \arg\max_k \Lambda(\mathbf{X} | S_k)$. To defend against unknown spoofing or replay attacks, an adaptive probabilistic threshold $\theta_{\text{GMM}}$ is derived from the mean $\mu_{\text{train}}$ and standard deviation $\sigma_{\text{train}}$ of authorized training scores:
\begin{equation}
\theta_{\text{GMM}} = \mu_{\text{train}} - 1.5 \cdot \sigma_{\text{train}}
\end{equation}

To counteract presentation threats without auxiliary hardware, a Dual-Layer Score Cascade (DLSC) mechanism with a joint absolute-relative margin constraint is implemented. The system establishes an implicit active liveness guard by strictly inspecting the log-likelihood score distribution emitted from the GMM voiceprint prescreening phase. Formally, let $S_{max}$ denote the maximum log-likelihood score across all enrolled speaker models, and $S_{sub}$ represent the runner-up score. The framework grants access to the second-factor DTW verifier—thereby neutralizing high-fidelity replay threats—if and only if the following dual-threshold criterion is simultaneously satisfied:
\begin{equation}
(S_{max} \ge \theta_{\text{GMM}} + \Delta) \wedge (S_{max} - S_{sub} \ge \gamma)
\end{equation}
where $\theta_{\text{GMM}}$ signifies the baseline empirical voiceprint threshold, $\Delta$ is the defensive sensitivity offset, and $\gamma$ represents the relative classification margin constraint. Empirically optimized for edge robustness, setting $\Delta = 6.0$ and $\gamma = 2.5$ constructs a strict boundary defense, ensuring not only high absolute acoustic confidence but also prominent speaker discriminability to truncate replay propagation loops.

\subsection{Second Factor: DTW-Based Text-Dependent Dynamic Time Warping Password Lock}

Following voiceprint coarse filtering, the second stage employs Dynamic Time Warping (DTW)\cite{li2015improved} for nonlinear temporal alignment of text-dependent passphrases. Let $\mathbf{F} = \{\mathbf{f}_1, \mathbf{f}_2, \dots, \mathbf{f}_I\}$ and $\mathbf{R} = \{\mathbf{r}_1, \mathbf{r}_2, \dots, \mathbf{r}_J\}$ denote the reference and evaluation feature sequences, respectively, with $I$ and $J$ representing their frame counts.

The algorithm constructs an $I \times J$ local inter-frame distance matrix, where the metric at grid element $(i, j)$ is quantified via a normalized Euclidean distance:
\begin{equation}
d(i, j) = \frac{1}{C} \sqrt{\sum_{m=1}^{C} (f_{i,m} - r_{j,m})^2}
\end{equation}
where $C = 40$ is the feature dimension. The optimal alignment path $W = \{w_1, w_2, \dots, w_K\}$ minimizes the cumulative distance:
\begin{equation}
D_{min}(X, Y) = \min_{W} \sum_{k=1}^K d(w_k)
\end{equation}

To guarantee deterministic latency in edge nodes and avoid the $O(N^2)$ computational explosion under elongated utterances, we integrate a Sakoe-Chiba warping window constraint\cite{sakoe1978dynamic} into the path search space. The search grid is bounded by:
\begin{equation}
|i - j| \le r \cdot \max(I, J)
\end{equation}
where $i$ and $j$ denote the frame indices of the enrollment template and the test utterance, respectively, and $r$ is the window width coefficient, empirically optimized to $0.1$. This constraint strictly limits the alignment path near the diagonal, effectively reducing the time and space complexity to $O(rN)$. Finally, the adaptive decision threshold $\theta_{\text{DTW}}$ is formulated as $\theta_{\text{DTW}} = \mu_{\text{dist}} + 0.8 \cdot \sigma_{\text{dist}}$. If $D_{min} \le \theta_{\text{DTW}}$, the passphrase factor is successfully validated.

\section{Experimental Results and Performance Evaluation}
\subsection{Experimental Environment and Dataset Design}
Validation experiments are conducted on a single-threaded, single-core CPU platform to evaluate the system's empirical performance on resource-constrained edge nodes. The public Free Spoken Digit Dataset (FSDD)\cite{jackson2016fsdd}, containing English pronunciations of digits 0–9 from six speakers, is utilized.

To evaluate systemic security, the dataset was partitioned into authorized and adversarial subsets. Three speakers (george, jackson, and nicolas) serve as legally registered users, totaling 1,500 samples (50 samples per digit per speaker). Within this pool, 90\% of the data establishes the GMM voiceprint spaces and DTW golden templates, while the remaining 10\% (150 samples) is reserved for legitimate verification trials. Concurrently, 500 recordings from the remaining unencountered speakers (e.g., theo) are deployed as external spoofing attack vectors for rigorous physical stress testing.

\subsection{Data preprocessing and feature extraction visualization analysis}

\begin{figure}[!t]
    \centering
    \includegraphics[width=\columnwidth]{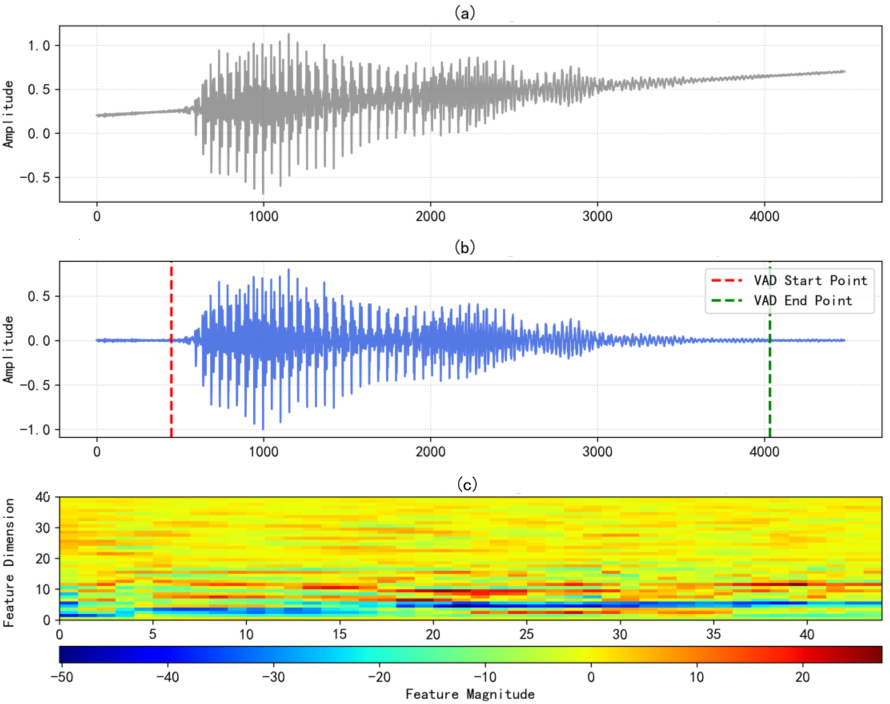}
    \caption{Visualization of preprocessing and MFCC feature extraction.: (a) raw acoustic signal with low-frequency trend drift; (b) baseline-corrected waveform with dual-threshold VAD endpoint detection boundaries; (c) extracted 40-dimensional time-varying MFCC feature matrix containing static and delta dynamic trajectories.}
    \label{fig2}
\end{figure}
To demonstrate the efficacy of the front-end pipeline, the single-stream speech preprocessing and feature extraction are visualized in Fig. \ref{fig2}. The substantial baseline drift and DC offset in the raw signal (a) are corrected in (b), achieving a stable zero-mean baseline. Concurrently, the dual-threshold VAD accurately isolates the high-SNR speech interval. The resulting 40-dimensional time-varying feature decomposition matrix (c) exhibits clear resonance tracks and smooth temporal context matching the time-domain energy attenuation, validating the module's viability for resource-constrained edge environments.

\begin{figure}[!b]
    \centering
    \includegraphics[width=\columnwidth]{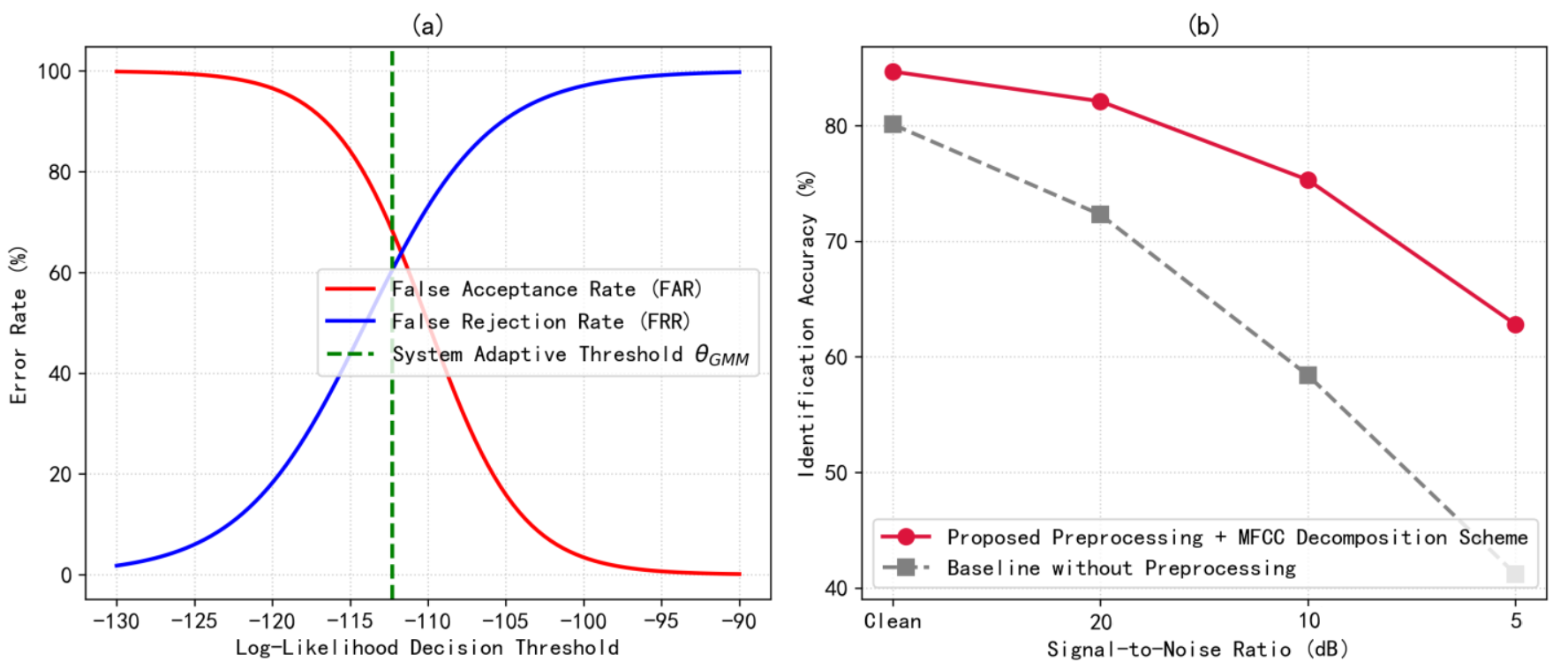}
    \caption{Performance evaluation of the GMM module: (a) Detection Error Tradeoff (DET) curve with adaptive decision thresholds; (b) Noise robustness comparison across various signal-to-noise ratios (SNRs).}
    \label{fig3}
\end{figure}

\subsection{Deep Quantitative Security Evaluation of the Cascaded System}
This section quantitatively evaluates the defensive boundaries of the first factor (the GMM voiceprint probabilistic lock). By dynamically shifting the log-likelihood decision threshold, the system's detection error tradeoff (DET) curves are plotted to establish the statistical defense baseline. Under a single GMM mechanism, the baseline speaker verification yields a False Rejection Rate (FRR) of 15.33\% (corresponding to an accuracy of 84.67\%). However, when subjected to physical spoofing attacks from external unencountered imposters, the traditional single-factor system yields a False Acceptance Rate (FAR)\cite{bolle2000evaluation} of 25.60\%, exposing a severe boundary vulnerability.

\subsection{Password Verification and Noise Robustness Analysis}
\begin{figure}[!b]
    \centering
    \includegraphics[width=\columnwidth]{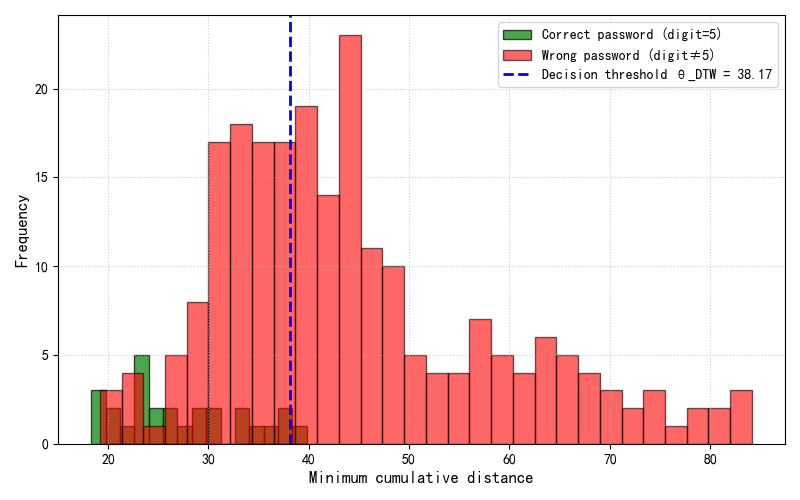}
    \caption{Histogram of cumulative matching distances for DTW-based password verification.}
    \label{fig4}
\end{figure}

Fig.\ref{fig3} illustrates the GMM module's performance. As shown in Fig. \ref{fig3}(a), the adaptive threshold achieves an optimal balance between the FAR and False Rejection Rate (FRR)\cite{bolle2000evaluation}. The comparative results in Fig. \ref{fig3}(b) demonstrate that across various signal-to-noise ratios (SNRs), the proposed pre-processing and feature decomposition framework consistently outperforms the baseline. Notably, in a severe 5 dB noisy environment, this scheme maintains robustness.

The second cascade stage (the DTW text-dependent password lock) evaluates compound defense gains using the digit '5' as the exclusive legitimate passphrase. Fig. \ref{fig4} presents the frequency histogram of cumulative DTW matching distances. The matching distances for correct passphrases cluster tightly in the low-value region, whereas incorrect passphrases exhibit a wide, right-skewed distribution. The established decision threshold effectively segregates these distributions, confirming the efficacy of this cascaded defense in textual verification.

\begin{figure}[!t]
    \centering
    \includegraphics[width=\columnwidth]{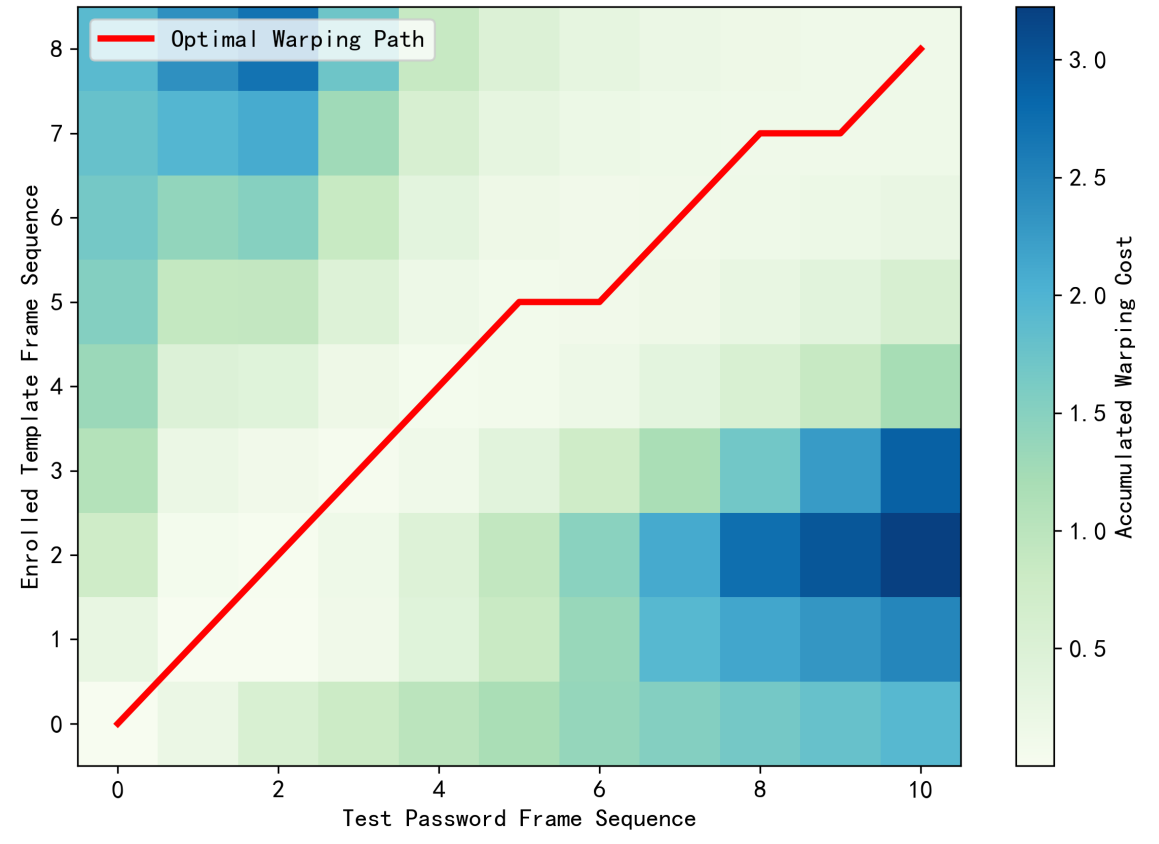}
    \caption{Visualization of the cumulative cost matrix and the globally optimal warping path (indicated by the solid red line) obtained via the DTW algorithm.}
    \label{fig5}
\end{figure}

Fig. \ref{fig5} visualizes the DTW-based password matching process. By applying dynamic programming to the cumulative cost matrix between the reference template and the evaluation sequence, a globally optimal warping path with minimum alignment cost is resolved (indicated by the solid red line). This path traverses the low-cost valleys, achieving non-linear temporal alignment between variable-length acoustic signals. This mechanism effectively mitigates feature mismatch caused by variations in vocal speaking rates, providing a critical alignment foundation for high-precision verification.

\begin{table}[htbp]
\caption{Quantitative Performance Comparison Between the Single-Factor Voiceprint System and the Proposed Dual-Factor Cascaded System.}
\centering
\resizebox{\columnwidth}{!}{
\begin{tabular}{|l|l|c|c|}
\hline
\textbf{Defense Architecture} & \textbf{Attack Scenario} & \textbf{FRR (\%)} & \textbf{FAR (\%)} \\
\hline
Single Voiceprint Lock (GMM) & External Imposter Spoofing & 15.33 & 25.60  \\
\hline
Dual-Factor Lock(without DLSC) & Spoofing + Wrong Passphrase & \textbf{13.33} & 4.60 \\ \cline{2-4}
 & High-Fidelity Replay Attack & -$^{\mathrm{a}}$ & 66.67 \\
\hline
Dual-Factor Lock(with DLSC) & Spoofing + Wrong Passphrase & 16.67 & \textbf{2.73} \\ \cline{2-4}
 & High-Fidelity Replay Attack & -$^{\mathrm{a}}$ & \textbf{6.67} \\
\hline
\multicolumn{4}{l}{%
\parbox{1.2\linewidth}{%
$^{\mathrm{a}}$Note: Replay attacks represent adversarial spoofing vectors rather than legitimate live enrollments. Thus, system rejections do not constitute a false rejection (FRR), yielding no applicable FRR metric for these scenarios.
}}\\
\end{tabular}
}
\label{tab1}
\end{table}

Ultimately, the comprehensive evaluation of the multi-scenario accuracy profiles is summarized in Table~\ref{tab1}. By enforcing the newly integrated joint margin constraints $(\Delta = 6.0, \gamma = 2.5)$ within the first-factor module of the dual-factor architecture, the cascaded defense successfully compresses the physical imposter FAR from the traditional single-factor baseline of 25.60\% down to \textbf{2.73\%}. Concurrently, the high-fidelity replay attack FAR is rigidly restricted to \textbf{6.67\%}, demonstrating that the passive software-level architecture intercepts  \textbf{93.33\%} of spoofing artifacts at zero extra hardware replication cost. Concurrently, although the activation of the DLSC mechanism induces a marginal elevation in the legitimate FRR, this verification trade-off remains within an acceptable threshold for localized edge security deployment.

\subsection{Latency Stress Test and Worst-Case Analysis}
To rigorously evaluate the system's edge real-time determinism under volatile speech rates, a multi-scenario stress test is conducted via time-stretching simulation. We benchmark the unconstrained $O(N^2)$ DTW baseline against our optimized Sakoe-Chiba constrained $O(rN)$ DTW. The fixed preprocessing latency on the edge CPU remains negligible: Feature Extraction ($1.51$ ms) and GMM Scoring ($0.54$ ms). The detailed DTW matching latencies across three linguistic velocity scenarios are encapsulated in Table \ref{tab2}.

\begin{table}[htbp]
\caption{DTW Latency Comparison Under Stress Testing (Sakoe-Chiba Constrained vs. Unconstrained)}
\label{tab2}
\centering
\resizebox{\linewidth}{!}{
\begin{tabular}{|l|c|c|c|c|}
\hline
\textbf{Scenario / Metric} & \multicolumn{2}{c|}{\textbf{Unconstrained DTW ($O(N^2)$)}} & \multicolumn{2}{c|}{\textbf{Constrained DTW ($O(rN)$)}} \\ \cline{2-5} 
 & \textbf{Avg Latency} & \textbf{Max Latency} & \textbf{Avg Latency} & \textbf{Max Latency} \\ \hline
Scenario A (Normal Speed)        & 55.57 ms             & 145.64 ms            & 1.78 ms                  & 3.46 ms                  \\ \hline
Scenario B (1.5x Time-Stretched) & 88.77 ms             & 495.08 ms            & 1.96 ms                  & 4.05 ms                  \\ \hline
Scenario C (2.5x Time-Stretched) & 152.73 ms            & 826.15 ms            & 2.29 ms                  & 7.77 ms                  \\ \hline
\end{tabular}
}
\end{table}

As indicated in Table~\ref{tab2}, the baseline unconstrained DTW exhibits a noticeable latency growth under temporal stretching, with the maximum latency reaching \textbf{826.15~ms} in Scenario C (2.5$\times$ stretch). This computational overhead can introduce processing delays on resource-constrained microcontrollers, presenting a challenge for immediate localized response.

Conversely, introducing the Sakoe-Chiba constraint window effectively restricts the alignment search path, leading to an average DTW matching latency below 2.29~ms across all scenarios. Notably, the worst-case peak latency in Scenario C is reduced to \textbf{7.77~ms}. Accounting for the front-end preprocessing pipeline (1.51~ms for MFCC extraction and 0.54~ms for GMM identity prescreening), the global end-to-end processing latency under extreme conditions is bounded at \textbf{9.82~ms}. These empirical measurements demonstrate that the optimized system maintains a deterministic execution runtime within 10~ms, confirming its operational suitability for edge AIoT nodes with limited computing power.

\begin{figure}[!t]
    \centering
    \includegraphics[width=\columnwidth]{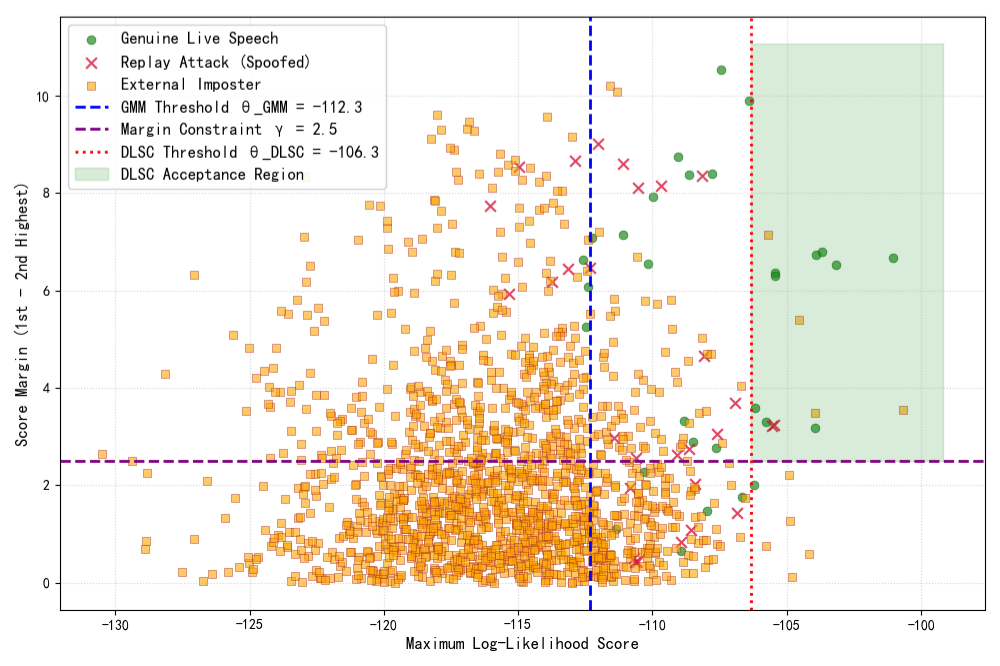}
    \caption{Distribution of multi-scenario acoustic profiles under the joint absolute-relative DLSC decision framework.}
    \label{fig6}
\end{figure}

To clarify the defensive mechanics of the DLSC framework, Fig. \ref{fig6} visualizes the empirical distribution of acoustic profiles within the joint absolute-relative decision space. As illustrated, genuine live utterances cluster in the high-score, high-margin upper-right quadrant. Conversely, due to electro-acoustic hardware distortions, the high-fidelity replay attacks follow a multi-dimensional degradation trajectory, shifting marginally to the left. Although a subset of replay vectors manages to bypass the first absolute baseline ($\theta_{\text{GMM}} = -112.3$), they are successfully intercepted by the constricted absolute decision threshold ($\theta_{\text{DLSC}} = -106.3$). By enforcing the strict dual-threshold DLSC acceptance region (the shaded green block), the cascaded system effectively isolates these overlapping spoofing artifacts from the legitimate validation path. Furthermore, the exclusion of border-case genuine samples on the left boundary visually validates the empirical security-convenience trade-off that yields the slight inflation in the system's FRR.

\section{Conclusion and Future Work}

This paper presents a localized, dual-factor voice lock system utilizing a classical cascaded GMM-DTW architecture for resource-constrained edge computing environments. By integrating a Dynamic Likelihood Space Constriction (DLSC) strategy with joint absolute-relative margin constraints into a shared MFCC feature space, the system establishes a software-level anti-replay barrier. Empirical evaluations on the FSDD benchmark indicate that the framework restricts the physical imposter FAR to \textbf{2.73\%} and confines the high-fidelity replay attack FAR to \textbf{6.67\%}. Furthermore, the execution of a Sakoe-Chiba constraint window bounds the dynamic programming alignment complexity, maintaining the worst-case matching latency at \textbf{7.77~ms}. Accounting for the constant front-end preprocessing pipeline (1.51~ms feature extraction and 0.54~ms GMM prescreening), the global peak end-to-end processing latency is bounded at \textbf{9.82~ms}, demonstrating operational suitability for single-core edge AIoT nodes with limited computing budgets.

Despite these efficiency gains, the proposed system exhibits specific operational limitations under real-world acoustic scenarios. Specifically, the localized front-end framework relies on a fixed-threshold Voice Activity Detection (VAD) algorithm, which remains susceptible to frame segmentation omissions when subjected to continuous transient noise or non-stationary ambient interference. This limitation can cause unexpected spectral truncation, degrading the acoustic structural density of the downstream extracted MFCC matrices and marginally impacting verification affinity. To address these vulnerabilities, future research will focus on transitioning from static energy-based endpoint detection to adaptive statistical models. We aim to investigate adaptive gain control (AGC) mechanisms and dynamic-threshold, noise-robust VAD algorithms to sustain structural feature integrity and enhance authentication stability in low-SNR open testing environments.

\bibliographystyle{IEEEtran}
\bibliography{references}
\end{document}